\pdfoutput=1
\documentclass[a4paper,american,floatfix,pdftex,superscriptaddress,twoside,%
aps,prl,
reprint,
]{revtex4-2}%
\usepackage{amsfonts,amsmath,amssymb}
\usepackage[T1]{fontenc}
\usepackage{graphicx}%
\usepackage[utf8]{inputenc}
\usepackage{mathtools}
\usepackage{hyperref, hypernat}
\usepackage[displaymath,textmath,graphics]{preview}

\usepackage{color}

\newcommand{\bra}[1]{\langle #1|}
\newcommand{\ket}[1]{|#1\rangle}

\newcommand{\COtwo}{$^{12}$C$^{16}$O$_2$}
\newcommand{\2}{$_2$}
\newcommand{\cm}{cm$^{-1}$}
\newcommand{\invcm}{cm$^{-1}$}
\newcommand{\ai}{\textit{ab initio}}
\newcommand{\um}{$\mu$m}

\newcommand{\grenoble}{\affiliation{Univ. Grenoble Alpes, CNRS, LIPhy, 38000 Grenoble, France}}%
\newcommand{\ucl}{\affiliation{Department of Physics and Astronomy, University College London, London, WC1E 6BT, UK}}%
\newcommand{\cfeldesy}{\affiliation{Center for Free-Electron Laser Science, Deutsches Elektronen-Synchrotron DESY, Notkestraße 85, 22607 Hamburg, Germany}}%
\newcommand{\uhhcui}{\affiliation{Center for Ultrafast Imaging, Universität Hamburg, Luruper Chaussee 149, 22761 Hamburg, Germany}}%
\newcommand{\uhhphys}{\affiliation{Department of Physics, Universität Hamburg, Luruper Chaussee 149, 22761 Hamburg, Germany}}%

\newcommand{\ayemail}{\email{andrey.yachmenev@cfel.de}}%
\newcommand{\acemail}{\email{alain.campargue@univ-grenoble-alpes.fr}}%
\newcommand{\syemail}{\email{s.yurchenko@ucl.ac.uk}}%

\begin{document}

\title{Electric quadrupole transitions in carbon dioxide}

\author{Andrey Yachmenev}\ayemail\cfeldesy\uhhcui%
\author{Alain Campargue}\acemail\grenoble
\author{Sergei N. Yurchenko}\syemail\ucl
\author{Jochen Küpper}\cfeldesy\uhhcui\uhhphys%
\author{Jonathan Tennyson}\ucl
\date{\today}%

\begin{abstract}\noindent%
Recent advances in the high sensitivity spectroscopy have made it possible, in combination with
accurate theoretical predictions, to observe for the first time very weak electric quadrupole
transitions in a polar polyatomic molecule of water.
Here we present accurate theoretical predictions of the complete quadrupole ro-vibrational spectrum
of a non-polar molecule CO\2, important in atmospheric and astrophysical applications.
Our predictions are validated by recent cavity enhanced absorption spectroscopy measurements and
are used to assign few weak features in the recent ExoMars ACS MIR spectroscopic observations
of the martian atmosphere.
Predicted quadrupole transitions appear in some of the mid-infrared CO\2 and water vapor
transparency regions, making them important for detection and characterization of the minor
absorbers in water- and CO\2-rich environments, such as present in the atmospheres of Earth, Venus
and Mars.
\end{abstract}

\maketitle

The intensities of electric quadrupole (E2) transitions are known to be very weak, six to eight
orders of magnitude smaller than the intensities of electric dipole (E1) transitions.
Until very recently, the E2 transitions were measured only in non-polar or slightly polar
diatomics, such as H\2~\cite{49Herzberg.H2, 81GoReRo.N2, 86ReJeBr.N2, 95BaMiLa.H2, 12CaKaPa.H2,
12HuPaCh.H2}, O\2 \cite{97NaLaUb.O2,09LoHaOk.O2}, N\2~\cite{44ReSiRo.O2, 81RoGoxx.O2, 12KaCaxx.CO2, 13KaGoCa.N2, 17CeVaMo.N2}, HD
\cite{16VaMoKa.H2}, N$_2^+$ \cite{14GeToWi.N2+}, i.e. molecules that otherwise do not exhibit E1
transitions or, as regards HD, they are extremely weak. In polar molecules, with polyatomics
being considerably richer in the number and density of ro-vibrational transitions, strong E1
absorption profiles blanket most of the weak features in the ro-vibrational spectrum.

Tracing and assigning weak spectral features as belonging to the E1 transitions of minor isotopologues or other meager molecular species, or indeed the E2 or magnetic dipole (M1) transitions of the main molecular constituent, can be
endlessly intricate and hence can benefit from precise theoretical predictions.
So far as there were no reliable calculations of the E2 and magnetic dipole M1 transitions for  polyatomic molecules, detection of these electric-dipole-forbidden features  and investigation of their seemingly surreptitious role in high-resolution spectroscopy remains unexplored.

The role of weak ro-vibrational transitions in spectrum of carbon dioxide (CO\2) is
fundamental to monitoring its isotopic composition in the atmosphere for understanding of carbon
cycle processes~\cite{83MoKoCa.CO2, 05Kerstel.CO2, 15Graven.CO2, 19LeKlPo.CO2} and it is becoming
increasingly precise for determination of new gas
signatures in CO\2-rich planetary atmospheres, such as Mars and Venus.
To this end, the completeness of spectroscopic data for the main isotopologue \COtwo\ is crucial, especially in its transparency windows, to reduce the likelihood of its weak spectroscopic features being mistakenly assigned to those of minor isotopologues or other less-abundant molecular species \cite{15McOgBe.CO2, 20TrPeKo.CO2}.
For example, recently discovered weak M1 transitions of CO\2\ in the spectrum of the martian atmosphere~\cite{20TrPeKo.CO2} would not be known as belonging to CO\2\ without the accurate knowledge of the E1 transitions of its main and minor isotopologues.
While there have been multiple and still ongoing computational and experimental efforts to fully
characterize the E1 spectra of major isotopologues of CO\2~\cite{jt678,jt700,jt804}, the contributions from the
dipole-forbidden E2 and, until very recently M1~\cite{20TrPeKo.CO2}, transitions have not had the slightest attention.

In this contribution, we report the first complete and accurate E2 line list for carbon dioxide \COtwo, which was validated in the 3.3 $\mu$m transparency window by recent laboratory measurements using Optical-Feedback-Cavity Enhanced Absorption Spectroscopy (OFCEAS)~\cite{21FlGrMo.CO2}. Based on our predictions and laboratory measurements, we report a detection of the E2
lines in the martian atmospheric spectra recorded by the ExoMars Trace Gas Orbiter Atmospheric Chemistry Suite (ACS) instrument~\cite{18KoBeDo}.
We also present our newly developed variational computational methodology, which is capable of  high-accuracy predictions of the E2 spectra for arbitrary polar molecules.

Our computational approach is based on the general variational approach TROVE \cite{TROVE, 15YaYuxx.method, 17YuYaOv.methods, jt730}, which for triatomic molecules employs an exact kinetic energy operator~\cite{20YuMexx}. For CO\2, an accurate empirically refined potential energy surface (PES) `Ames-2' was employed~\cite{17HuScFr.CO2}. In TROVE, the vibrational basis set is constructed in a multi-step procedure from contracted and symmetry-adapted products of one-dimensional basis sets, each represented by solutions $\chi_n(q)$  ($n=0,1,2...$) of the one-dimensional Schr\"{o}dinger equation for a selected vibrational mode $q$~\cite{17YuYaOv.methods}. For CO\2 there are two stretching and one bending vibrational modes, with the respective quantum numbers $n_1$, $n_2$, $n_3$. Since CO\2 is linear, its bending vibration is coupled to a molecular rotational motion about the linearity axis, which is the molecular $z$ axis. For this reason, the bending basis functions cannot be fully decoupled from the molecular rotation and parametrically depend on the rotational quantum number $K$ of the $\hat{J}_z$ angular momentum operator, hence the notation $n_3^{(K)}$. We refer for details of treatment of linear and quasi-linear molecules to \cite{20YuMexx}. The size of vibrational basis was controlled by the condition $2(n_1 + n_2) + n_3^{(K)} \le  64$.

The full ro-vibrational basis set is constructed as a symmetrized product
of the symmetry-adapted vibrational basis functions $\psi_{\lambda,K}^{(\Gamma_{\rm
vib})}$ and symmetry-adapted rigid-rotor wave functions $\ket{J,K,\Gamma_{\rm rot}}$:
\begin{equation}
\Psi_{\lambda,K}^{(J,\Gamma)} = \{ \psi_{\lambda,K}^{(\Gamma_{\rm vib})} \times \,
\ket{J,K,\Gamma_{\rm rot}} \}^{\Gamma}.
\end{equation}
Here, $\lambda$ denotes a set of vibrational state quantum numbers, and
$\Gamma_\text{vib}$, $\Gamma_\text{rot}$, and $\Gamma$ denote the symmetries of the vibrational, rotational and total wave functions, respectively.
For CO\2\ we employed the $\mathbf{C}_\text{2v}\text{(M)}$ molecular symmetry group. The total wavefunction for a ro-vibrational state $l$, with the quantum number of the total angular  momentum $J$ and the total symmetry $\Gamma$, is a linear combination of ro-vibrational basis set functions:
\begin{equation}
\Phi_{l}^{(J,\Gamma)} = \sum_{K,\lambda}c_{K,\lambda}^{(J,\Gamma,l)}\Psi_{\lambda,K}^{(J,\Gamma)},
\end{equation}
where the linear expansion coefficients $c_{K,\lambda}^{(J,\Gamma,l)}$ are obtained by solving
an eigenvalue problem with the full ro-vibrational Hamiltonian.
All energies and eigenfunctions up to $J=40$ were generated and used to produce the E2 line
list for CO\2.

The achieved accuracy of energy level predictions for CO\2 is best characterized by the
root-mean-squares
(rms) deviation of 0.06~\cm\ between the calculated and experimental~\cite{jt691s} ro-vibrational
term values, evaluated across 337 band centers with energies up to $15\,500$~\invcm\ above
zero-point level.
Considering the weakness of the E2 lines, the line position accuracy is crucial for discriminating
them from the weak E1 lines of minor isotopologues and possible impurities.
Further improvement of accuracy would require yet another round of empirical adjustment of the
underlying PES, which is an inordinately expensive procedure.
As a call for more practical solution very often the variationally computed energies, providing
they are
close enough to experiment, are replaced by  the corresponding experimentally determined values.
The latter can be extracted from the experimental spectroscopic line positions using advanced
combination difference techniques such as, for example, MARVEL (Measured Active
Rotational-Vibrational Energy
Levels)~\cite{jt412, 07CsCzFu.method, 12FuCsxx.methods, jt750} and RITZ (Rydberg-Ritz combination
principle) \cite{09MiTaPu.H2O, 10TaVeMi.CO}.
For CO\2, we used the experimental energy levels from the Carbon Dioxide Spectroscopic
Databank~\cite{15TaPeGa.CO2} produced from a global effective Hamiltonian modeling
of an exhaustive set of position measurements available in the literature~\cite{98TaPeTe.CO2}.
More details about the computational procedure  employed here and validation of its accuracy
are presented in a study of the E1 line list of CO\2~\cite{jt804}.

The quadrupole spectrum was simulated using the variational approach RichMol~\cite{RichMol,
RichMol_2}, a computer program designed for calculations of molecular ro-vibrational dynamics in
the presence of an external electromagnetic field. The transition probability from an initial ro-vibrational state $|i\rangle = \Phi_{l}^{(J,\Gamma)}$
into a final state $|f\rangle = \Phi_{l'}^{(J',\Gamma')}$ due to the interaction of light with
quadrupole moment of molecule is given by
\begin{eqnarray}
P(f\leftarrow i) = g_\text{ns}(2J'+1)(2J+1)
\left|\mathcal{K}^{(J',\Gamma',l',J,\Gamma,l)} \right|^2
\end{eqnarray}
where
\begin{align}\label{eq:kmat}
\mathcal{K}^{(J',\Gamma',l',J,\Gamma,l)} &= \sum_{K'\lambda'}\sum_{K\lambda}
\left[c_{K',\lambda'}^{(J',\Gamma',l')}\right]^* c_{K,\lambda}^{(J,\Gamma,l)}  \\ \nonumber
&\times (-1)^{K'} \sum_{\sigma=-2}^{2}
\left(\begin{array}{ccc}J&2&J'\\K&\sigma&-K'\end{array}\right) \\ \nonumber
&\times \sum_{\alpha,\beta=x,y,z} U_{\sigma,\alpha\beta}\bra{\psi_{\lambda',K'}}
Q_{\alpha,\beta}\ket{\psi_{\lambda,K}}.
\end{align}
Here, $Q_{\alpha,\beta}$ denotes the traceless quadrupole moment tensor in
the molecular frame and the matrix $U_{\sigma,\alpha\beta}$ transforms quadrupole tensor from
Cartesian to spherical-tensor form (see, e.g., Eqs. (5-41)--(5-44) in \cite{88Zare.book}).
The nuclear spin statistical factors $g_\text{ns}$ for \COtwo\ are equal to one for initial state
symmetries $A_1$ and $A_2$ and zero otherwise.

The quadrupole moment tensor with elements as functions of internal coordinates is required to
compute the vibrational expectation values $\bra{\psi_{\lambda',K'}}
Q_{\alpha,\beta}\ket{\psi_{\lambda,K}}$ in
Eq.~\eqref{eq:kmat}. A three-dimensional mesh of internal coordinates of CO\2 was used, containing
about 2000 different nuclear geometries and covering the energy range up to $hc\cdot$40\,000~\invcm\
above the equilibrium energy. The quadrupole tensor at each point was computed using all-electron
CCSD(T) level of theory in conjunction with the aug-cc-pwCVQZ basis set~\cite{89Dunning.ai,
92KeDuHa.ai, 02PeDuxx.ai} and analytic gradient approach~\cite{91Scuseria.ai}.
The electronic structure calculations employed the quantum chemistry package CFOUR~\cite{CFOUR}.
For a molecular frame selected such that the $x$ axis bisects the valence bond angle O--C--O and
plane of the molecule for bent CO\2 is aligned with the $xz$ plane, the three non-zero elements
of the quadrupole tensor $Q_{xx}$, $Q_{zz}$ and $Q_{xz}$ have the $A_1$, $A_1$, and $B_2$
symmetries, respectively. These were parameterized using fourth
order symmetry-adapted power series expansions through least-squares fittings to the electronic
structure data, with $\sigma_\text{rms}<10^{-4}$ a.u.



Our results for the quadrupole moment of CO\2 $Q_{zz}^\text{e} = -3.1666$~a.u. = $-14.207\times 10^{-40}$~Cm$^2$ at the equilibrium $r_\text{e}$ = 1.1614~\AA\ and its zero-point vibrational average $Q_{zz}^\text{ZPVA} = -3.1627$~a.u. = $-14.190\times 10^{-40}$~Cm$^2$ agree very well with experimental data and previous calculations listed in Table~\ref{t:quadr}.

\begin{table}
\caption{Experimental and theoretical permanent electric quadrupole moments of CO\2\ in $10^{-40}$ Cm$^2$}
\label{t:quadr}
\begin{tabular}{lrc}
\hline
Method & Value & Source \\
\hline
\emph{Experiment} \\
Buckingham effect\footnotemark[1] &$-14.98 \pm 0.50$  & \citep{81BaBuNe.CO2} \\
Buckingham effect\footnotemark[1] &$-14.3 \pm 0.6 $   & \citep{97WaCrRi.CO2} \\
Buckingham effect\footnotemark[1] &$-14.31 \pm 0.74 $  & \citep{11ChCoxx.CO2} \\
Cotton-Mouton effect\footnotemark[2] &$-14.0 \pm 0.7 $  & \cite{84KlHuxx.CO2} \\
Collision-induced absorption 
&$-14.9\pm 0.7$     & \citep{71HoBiRo.CO2} \\
Dielectric measurements &$-14.94 \pm 1.0 $  & \citep{97HoStBo.CO2} \\
\emph{Theory} \\
CCSD(T)/CBS\footnotemark[3] & $-14.22 \pm 0.09$ & \citep{00CoHaRi.CO2} \\
CCSD(T)/CBS + ZPVC\footnotemark[4] & $-14.29\pm 0.09$  & \citep{00CoHaRi.CO2} \\
CCSD(T)/6s4p4d1f & $-14.3$ &  \citep{03Maroulis.CO2} \\
CCSD(T)/AwCVQZ & $-14.207$ & This work \\
CCSD(T)/AwCVQZ/ZPVC\footnotemark[4] & $-14.190$ & This work \\
\hline
\end{tabular}
\footnotetext[1]{Electric field gradient induced birefringence measurements.}
\footnotetext[2]{Magnetic field induced birefringence measurements.}
\footnotetext[3]{Complete basis set limit extrapolation with additive core-valence correlation effect.}
\footnotetext[4]{Results corrected with the  zero-point vibration.}
\end{table}

The line intensity of quadrupole transition in units cm/molecule is given by
\begin{eqnarray}
I(f\leftarrow i) = \frac{4\pi^5 \nu^3
e^{-\beta E_l^{(J,\Gamma)}}\left(1-e^{-\beta hc\nu}\right)}{(4\pi\epsilon_0)5hcZ(T)}P(f\leftarrow
 i),
\end{eqnarray}
where $\nu=(E_{l'}^{(J',\Gamma')}-E_l^{(J,\Gamma)})/hc$ is the frequency of transition between
lower $E_l^{(J,\Gamma)}$ and upper $E_{l'}^{(J',\Gamma')}$ state energies, $Z(T)$ is the partition
function, and $\beta=1/(kT)$.
For \COtwo\ we used an accurate computed value of $Z(296~\text{K})=286.094$~\cite{jt678}.
The computed quadrupole line list is provided as supplementary material to the paper.

\begin{figure}
\includegraphics[width=\linewidth]{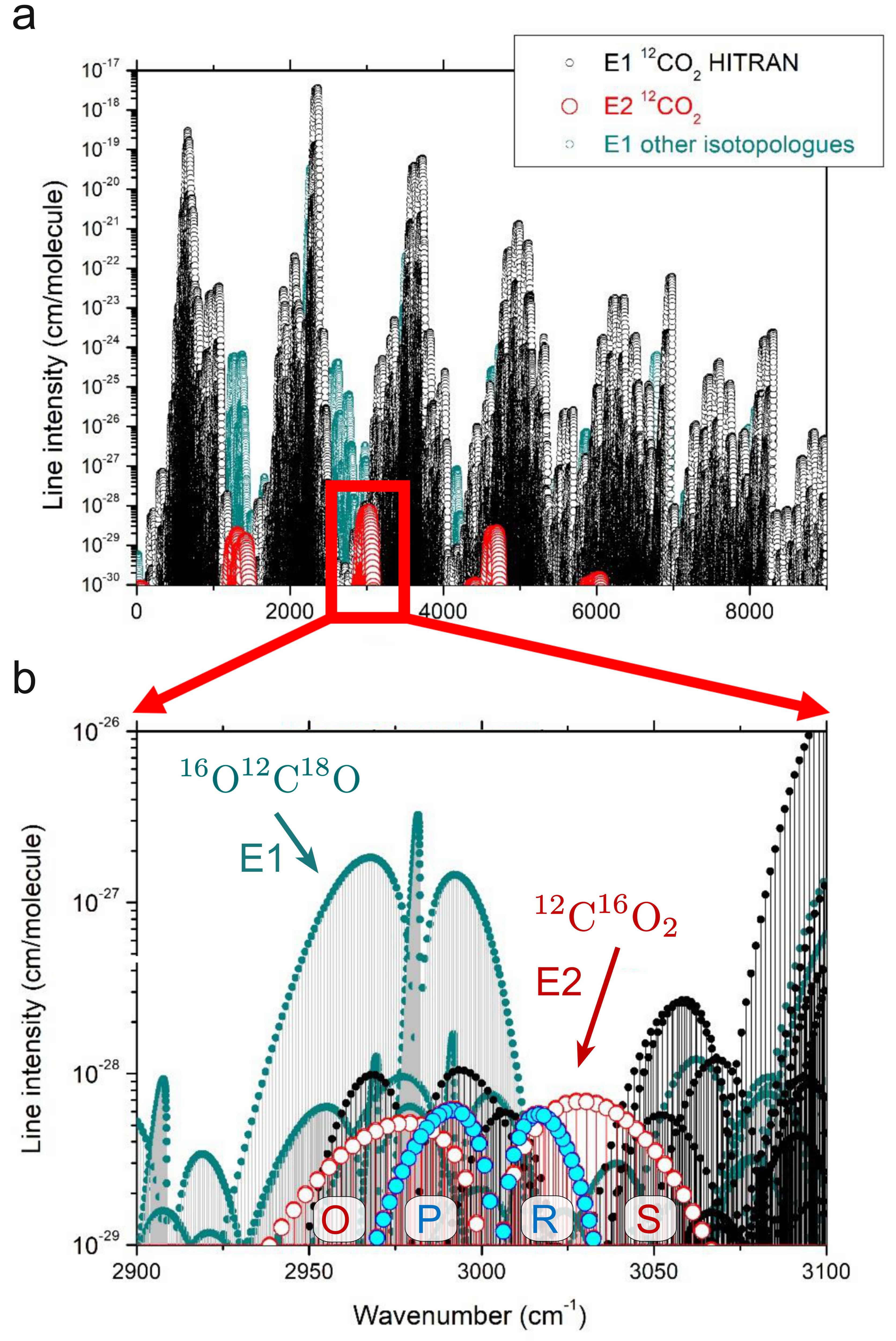}
\caption{(a) Overview of the calculated quadrupole spectrum of \COtwo (red) superimposed with its dipole spectrum (black) and the dipole spectrum of minor isotopologues (cyan). The spectra are generated for temperature $T=296$~K. (b) A zoom into the $\nu_2+\nu_3$ band, where the E2 transitions of $^{12}$C$^{16}$O$_2$ can be distinguished by larger circles, the \emph{P} and \emph{R} branches plotted with cyan colour and \emph{O} and \emph{S} branches plotted with red colour.}
\label{fig:CO2:E2:all}
\end{figure}

On the upper panel of Fig.~\ref{fig:CO2:E2:all}, the calculated E2 line list for the main
isotopologue, \COtwo, is superimposed on the E1 line list of the natural CO\2~\cite{jt691s}. Both
spectra are generated at room temperature, $T$ = 296 K. The E1 lines of the minor isotopologues at
natural abundances are plotted on the figure with cyan color. In general, the E2 bands are
6--8 orders of magnitude weaker than the E1 bands. The most prominent E2 band with line intensities of a few $10^{-29}$ cm/molecule is the bending plus anti-symmetric stretching
$\nu_2+\nu_3$ band at 3000 \cm\ (3.3 \um), highlighted on the lower panel of Fig.~\ref{fig:CO2:E2:all}.

In contrast with E2 spectrum of water~\cite{20CaKaYa.H2O, 20CaSoSo}, the E2 lines of CO\2 appear largely in the E1 transparency regions, which makes them possible to observe.
The reason for this is different approximate selection rules for the E1 and E2 transitions in CO\2
and in linear molecules in general. For example, the $\nu_1$ and $\nu_2+\nu_3$ E1 band transitions at 6.9 and 3.3~$\mu$m are dipole forbidden and therefore very weak. These bands  are however directly allowed for the E2 transition mechanism, even in the rigid-rotor approximation.

As already mentioned, the E1 lines of the minor isotopologues of CO\2 make an important
contribution to
the transparency windows (see Fig.~\ref{fig:CO2:E2:all}), where they overlap with the E2 lines of
the main isotopologue.
Indeed, the predicted $\nu_2+\nu_3$ E2 band of \COtwo\ is almost entirely superimposed with the
$\nu_2+\nu_3$ E1 band of $^{16}$O$^{12}$C$^{18}$O, which is stronger by about a factor 20, in spite
a natural isotopic abundance of only $4\times 10^{-3}$. It should be noted that due to the
different symmetry, the $\nu_2+\nu_3$ E1 band is allowed in $^{16}$O$^{12}$C$^{18}$O but forbidden
in \COtwo.

\begin{figure}
\centering
\includegraphics[width=\linewidth]{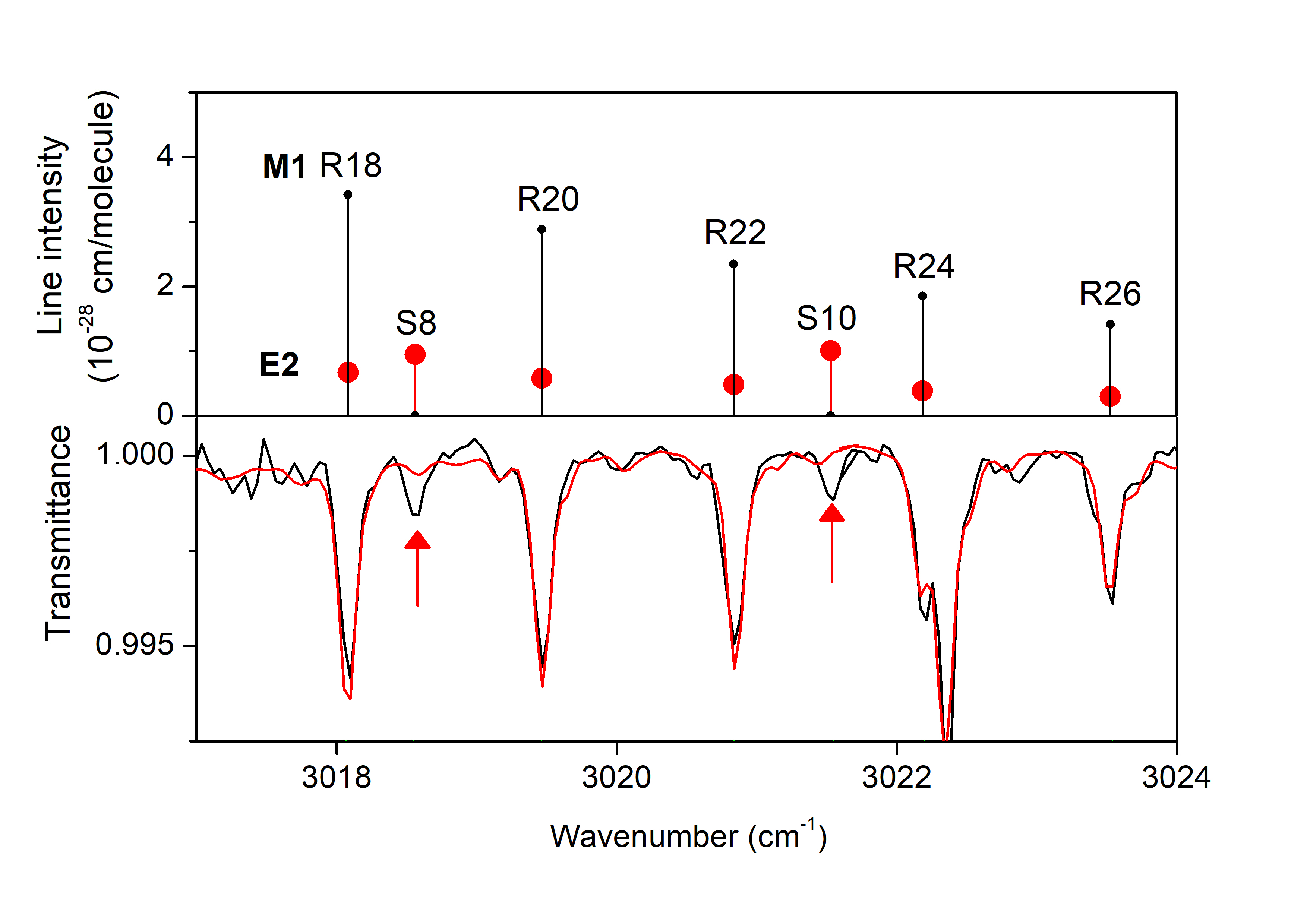}

\caption{ Comparison of the ACS MIR transmission spectra of the atmosphere of Mars~\cite{20TrPeKo.CO2} near 3020 \invcm\ with calculated line lists of the M1 and E2 $\nu_2+\nu_3$
bands of $^{12}$C$^{16}$O$_2$. Upper panel: The presently calculated E2 transitions with \ai\
intensities (red stems) are superimposed to the M1 transitions with intensities calculated
in \cite{21BoSoSo.CO2} (black stems).  Lower panel: ACS MIR spectrum (red line) with a best-fit
synthetic model containing the contributions of the E1 bands of CO\2\ and H\2O based on HITRAN 2016 database and of the M1 lines identified in \cite{20TrPeKo.CO2}. The assignments of the newly detected S8 and S10 E2 transitions are marked with red arrows.}
\label{fig:CO2:Mars}
\end{figure}

\begin{figure}
	\includegraphics[width=\linewidth]{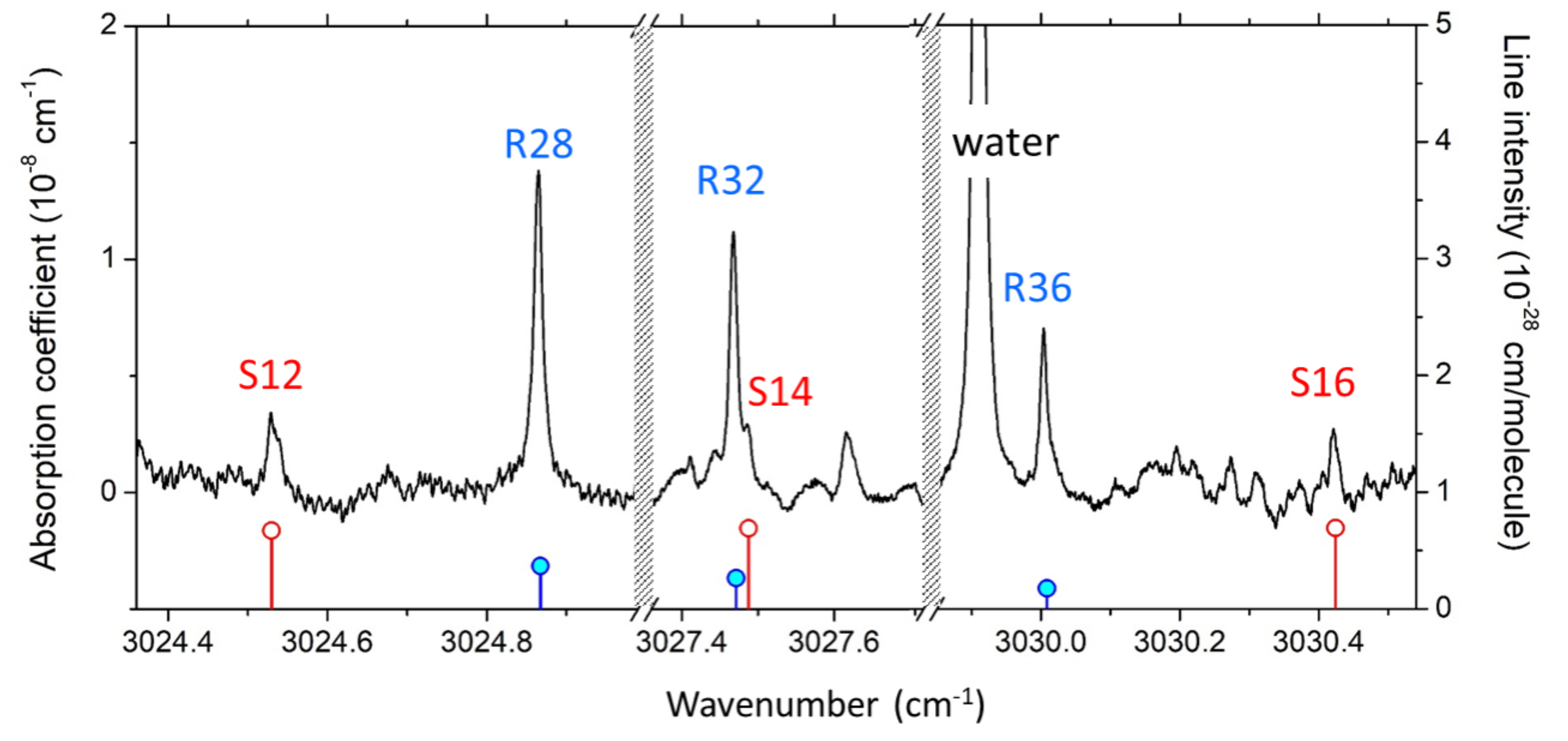}
	\caption{Comparison of the laboratory spectrum of CO\2 at 60~mbar recorded by OFCEAS~\cite{21FlGrMo.CO2} with the calculated E2 line list of \COtwo\
	in three spectral intervals of the E2 $\nu_2+\nu_3$ band.
	The $R$- and $S$-band E2 transitions are plotted with blue and red stems, respectively. The
	S12, S14 and S16 E2 lines are apparent. The line intensities of the R28,
	R32 and R36 transitions is mostly due to the M1 transitions.}
	\label{fig:co2_ofceas}
\end{figure}

The $\nu_2+\nu_3$ band of CO\2 has recently been observed in the spectrum of Mars' atmosphere~\cite{20TrPeKo.CO2}, as recorded by the ExoMars
Trace Gas Orbiter ACS-MIR (Atmospheric Chemistry Suite Mid InfraRed) spectrometer. The recorded spectra (partly reproduced in Fig.~\ref{fig:co2_ofceas}) show strong intensity features that are more specific for the M1 rather than the E2 transitions \cite{21PeTrLu.CO2}, although no first principles simulations for the M1 spectrum had been done.
In particular, the presence of the strong $Q$-branch, which is very weak for the E2 band, see
Fig.~\ref{fig:CO2:E2:all}(b), and the absence of nearly as intense the $O$- and $S$-branches argue for a
dominant M1 mechanism \cite{21PeTrLu.CO2}.
But this does not exclude presence of the E2 features in the martian spectrum, although obviously
they must be less pronounced.
Note that very recently, the dominant M1 $\nu_2+\nu_3$ band was also detected in the laboratory from long path absorption measurements of \COtwo\ by Fourier transform spectroscopy (FTS) \cite{21BoSoSo.CO2}. These measurements indicated that the M1 line intensities estimated from martian spectrum \cite{20TrPeKo.CO2} were overestimated by about a factor of two. The effective dipole moment derived from a fit of the measured FTS intensities was used to calculated the line intensities displayed in Fig. \ref{fig:CO2:Mars}.

As can be seen from Fig.~\ref{fig:CO2:E2:all}(b), E2 transitions in the $O$-branch
overlap completely with much stronger and generally broader E1 lines of CO\2, and hence are hardly
discernible. A number of the $S$-branch transitions appear in a narrow region free from any strong
E1 lines, which are thus possible to observe.
In Fig.~\ref{fig:CO2:Mars}, we compare our predicted $S$-branch transitions near 3020~\invcm\ to
the ACS MIR transmission spectra displayed on Fig.~4 of \cite{20TrPeKo.CO2}.
A clear position coincidence of the predicted $S$(8) and $S$(12) lines with
unknown weak absorption features can be observed in the region 3018--3025~\invcm.

A very recent laboratory study, dedicated to measurements of the weak spectral features in the $S$-branch of $\nu_2+\nu_3$ band of CO\2 by OFCEAS~\cite{21FlGrMo.CO2}, has confirmed the strong M1 features originally observed in martian CO\2. The measurements have also revealed a number of new features that can be assigned, by comparison with theoretical line list, to the S12, S14, and S16 E2 lines, shown in Fig.~\ref{fig:co2_ofceas}. Notably, the intensities of the E2 $R$-branch transitions contribute to about 10-15\% of those of the dominant M1 features.

Intensity considerations provide further evidence supporting the detection of E2 lines both in ACS MIR OFCEAS spectra. The intensity of the S8 and S10 E2 lines can be estimated from the martian spectrum by comparison to the nearby R18-R22 for which line intensities are known from FTS laboratory spectra of Ref.~\cite{21BoSoSo.CO2}. From a multi-line profile fit of the ACS MIR spectrum, the areas of the S8 and S10 lines were derived and scaled according to the absolute intensities of the nearby R18-R22 M1 lines \cite{21BoSoSo.CO2}. We obtain 7.0 and 6.5 $\times 10^{-29}$ cm/molecule at 172 K for the absolute intensities of S8 and S10 which is in satisfactory agreement with our theoretically predicted values of 9.5 and 10.0 $\times 10^{-29}$ cm/molecule, respectively. As concerns the S12, S14 and S16 lines observed in the OFCEAS spectra (Fig. \ref{fig:co2_ofceas}), their intensities (and pressure broadening coefficients) were reported in Ref. \cite{21FlGrMo.CO2}. The OFCEAS intensities are also close to their predicted values, although showing a difference on the order of 30\%, slightly larger than the estimated OFCEAS uncertainty.

The experience of \ai\ E2 intensity calculations  for polyatomic molecules is very limited. Based on the wealth of the E1 experimental and theoretical investigations, it is nowadays normal to expect a 10-20\% error (for stronger bands) from  \ai\ predictions of electric dipole intensities using standard levels of theory, as in this work (CCSD(T)/aug-cc-pwCVQZ). However the electric quadrupole calculations are still unexplored territory. According to the current work, our prediction of the two E2 lines from $\nu_2+\nu_3$ amounts to a 30\% (overestimated) error in intensities at $T=172$~K. In  fact, a similar quality of the \ai\ intensities  was reported for for the $\nu_1+\nu_3$ E2 transitions of water in \cite{20CaKaYa.H2O}, which varied between 8 and 90\% underestimating the experimental intensities. More work is needed, both experimental and theoretical, to establish the quality of the modern \ai\ methods for the E2 intensities.

In summary, we presented an accurate computational methodology for calculating the electric quadrupole spectra of polyatomic molecules with arbitrary structure. Calculated quadrupole transitions of CO\2\ were confirmed by the high sensitivity spectroscopic measurements~\cite{21FlGrMo.CO2} with few of them newly identified in the
spectrum of Mars' atmosphere.The quadrupole transitions are typically a million times weaker than the electric dipole  transitions.
The accurate characterization of the quadrupole transitions for the main atmospheric absorbers, especially in the mid-infrared transparency windows, will eliminate the misassignment of spectral features and thus help in precise detection of the minor atmospheric constituents. Being particularly sensitive to steeply varying fields, which are common in nature at the
molecule-molecule and molecule-surface interfaces~\cite{18KaKoBa.O2, 19RuStGl, 19MuJuxx}, the  electric quadrupole transitions can potentially be used for remote sensing of local molecular environments.

\section*{Supplementary material}

Contains a quadrupole line list for CO\2 in the HITRAN-like format computed at $T=296$~K with the threshold of $10^{-36}$ cm/molecule for the absorption coefficient.  The ro-vibrational states were assign using the HITRAN quantum-numbers convention for CO\2\ (see \cite{20YuMeFr}). The ExoMol diet \cite{jt684} scheme of CO\2\ was used for the broadening coefficients (air and self), while the line shifts were set to zero.

\section*{Acknowledgments}
We thank  Trokhimovskiy for providing the original ACS MIR transmission spectra of the atmosphere of Mars.
\\
This work was supported by the STFC Project No. ST/R000476/1 and ERC Advance Grant Project No. 883830. The authors acknowledge the use of the UCL Legion High Performance Computing Facility (Legion@UCL) and associated support services in the completion of this work, along with the Cambridge Service for Data Driven Discovery (CSD3), part of which is operated by the University of Cambridge Research Computing on behalf of the STFC DiRAC HPC Facility (www.dirac.ac.uk). The DiRAC component of CSD3 was funded by BEIS capital funding via STFC capital grants ST/P002307/1 and ST/R002452/1 and STFC operations grant ST/R00689X/1. DiRAC is part of the National e-Infrastructure. The work was further supported by the Deutsche Forschungsgemeinschaft (DFG) through the cluster of excellence ``Advanced Imaging of Matter'' (AIM, EXC~2056, ID~390715994) and through the Maxwell computational resources operated at Deutsches Elektronen-Synchrotron DESY, Hamburg, Germany.

\section*{Data Availability}
The data that supports the findings of this study are available within the article and its supplementary material.


%

\end{document}